# Polarization Origin of Photoconductivity in $MAPbI_3$ Thin Films

Rohit Saraf,[1] Cecile Saguy,[2] Vivek Maheshwari,[1] Hemaprabha Elangovan[2,3] and Yachin Ivry[2,3,*]

[1]Department of Chemistry, Waterloo Institute of Nanotechnology, University of Waterloo, Waterloo, Ontario N2L 3G1, Canada

[2] Solid State Institute, Technion – Israel Institute of Technology, Haifa 32000, Israel

[3]Department of Material Sciences and Engineering, Technion – Israel Institute of Technology, Haifa 32000, Israel

**Abstract:**

**Hybrid-halide perovskite (HHP) films exhibit exceptional photo-electric properties. These materials are utilized for highly efficient solar cells and photoconductive technologies. Both ion migration and polarization have been proposed as the source of enhanced photoelectric activity, but the exact origin of these advantageous device properties has remained elusive. Here, we combined microscale and device-scale characterization to demonstrate that polarization-assisted conductivity governs photoconductivity in thin HHP films. Conductive atomic force microscopy under light and variable temperature conditions showed that the photocurrent is directional and is suppressed at the tetragonal-to-cubic transformation. It was revealed that polarization-based conductivity is enhanced by light, whereas dark conductivity is dominated by non-directional ion migration, as was confirmed by large-scale device measurements. Following the non-volatile memory nature of polarization domains, photoconductive memristive behavior was demonstrated. Understanding the origin of photoelectric activity in HHP allows designing devices with enhanced functionality and lays the grounds for photoelectric memristive devices.**

Hybrid-halide perovskites (HHP), specifically methylammonium lead iodide ($MAPbI_3$), garner much interest in recent years thanks to their unique electro-optic and photo-electric properties. These materials are therefore utilized for competitive technologies, such as high efficiency solar cells, as well as self-powered photodetectors and devices.[1–5] The typical light-assisted electron-hole pairing and separation in semiconductors[6] might not suffice to support the observed photoconductivity in HHPs. Hence, additional mechanisms have been proposed as governing the photoconductivity in these materials. Traditionally, the advantageous properties of HHP devices are attributed to ion migration, which is a result of the high ion mobility in these semiconducting materials.[7,8] Recent accumulated literature discusses the existence of polarization and polarization domains in these materials**,** especially in highly strained geometries, such as thin films.[9–16] It has been proposed that the high photocurrents stem from polarization-based conductivity, which is a competing mechanism for ion migration.[17] Hence, it is currently uncertain what mechanism governs the advantageous material and device properties.

Upon electric-field poling, both ion migration and polarization effects are observable.[3,18,19] Both mechanisms can also support previously reported anomalous photovoltaic behavior, such as high photo-voltage and a hysteretic current-voltage curve is also supported by.[19,20] This causes significant challenges in decoupling the two effects, hence hindering the realization of the fundamental mechanism that governs the functionality in these materials, as well as encumbering their technological development. While polarization effects provide more self-powered and stable capacity for devices,[11,19] ion migration is transient in nature, leading to material degradation and shorter device lifetime.[18,20,21] Therefore, a basic understanding of the two processes at a nanoscale level, *i.e.*, at the domain and grain scale, is required. Discriminating between these two competing effects will significantly advance the ability to improve the relevant technologies and develop stable self-powered devices, *e.g.*, by prioritizing polarization effects over ion migration.

Distinguishing between ion migration and polarization-mediated conductivity is presumably possible by observing the exact paths of the dark- and photo-current in the material. In particular, while current related to ion migration is nondirectional and follows the shortest path, polarization-based conductivity is orientational, following the polarization orientation even if the path is longer. In addition, although both mechanisms prioritize conductivity along the grain boundaries, polarization conductivity enables also intra-grain conductivity through the domain walls .[22–25]

In this work, nanoscale charge transport characterization and macroscale device measurements were performed to discriminate between ion-migration and polarization-mediated photoconductivity, as well as dark conductivity, by means of direct observations. Conductive atomic force microscopy (c-AFM) mapping with *in-situ* temperature and illumination control was combined with macro-scale photoelectric measurements to demonstrate that orientational polarization-mediated conductivity is the principal mechanism behind the enhancement of photocarriers in HHP. Based on these findings, the existence of a memory effect was shown by measuring after-biasing non-volatile photoconductivity. The disappearance of photoconductive activity above the tetragonal-to-cubic transition is illustrated with the variable-temperature c-AFM in agreement with independent variable-temperature X-ray structural profiling. We showed that under dark conditions, ion migration is dominant, whereas polarization-based conductance is enhanced under the light. Moreover, it was demonstrated that above a threshold bias, grain boundaries and domain walls conduct better than the grains themselves. Likewise, photocurrent enhancement was demonstrated by introducing a non-polar polymer which hinders the pathway for ion migration. Good agreement between macroscopic and microscopic results was also demonstrated. Lastly, based on the presented memory effect, we propose a non-volatile photoconductive memory device.

**Results**

We present the results of various experiments designed to determine (1) the mechanism dominating the current both in light and dark, and (2) the effect of adding a non-FE polymer, polystyrene (PS) to plain $MAPbI_3$ to the photocurrent.

We first demonstrated that the mechanism dominating the photocurrent is based on orientational polarization and not on ion migration. c-AFM scan (Figure 1a) and device-scale solar simulator (Figure 1b) were used for multiscale current-flow characterization. Because ion migration is not directional, the shorter the path to the electrode, the higher is the output current. Contrariwise, for polarization conductance dominancy, one expects larger photocurrent flow along the polarization orientation. Thus, characterizing samples with a predominant in-plane orientation of the long tetragonal axis (*i.e.*, in-plane polarization, see X-ray diffraction [XRD] data in Figure S1) is expected to result in an in-plane polarization-assisted photoconductivity. Using the geometry of Figure 1a (see Methods for complementary material properties and device design), stronger currents through the 200 nm film thickness above the electrodes (out-of-plane) reflect ion migration. Similarly, stronger in-plane currents within the 2 μm wide channel express polarization-assisted photoconductivity dominancy.

Figure 1c shows the current distribution under light between the scanning tip and the device electrodes for $V_{bias}$ = 2 V for a 1 weight % PS-$MAPbI_3$ film, following the setup of Figure 1a. The scanned area comprises both out-of-plane current flow between the electrode and the tip through the 200 nm thick films, as well as in-plane flow across the 2 μm wide channel (the simultaneously imaged topography is given in Figure S3a). Figure 1c shows that despite being of a longer path, the in-plane photocurrent measured in the channel is higher than the out-of-plane current above the electrodes by almost two orders of magnitude. To demonstrate polarization-assisted photoconductivity dominancy within the channel as well as the overall enhanced grain boundary conductivity, we characterized areas within the channel and above the electrodes with c-AFM at higher magnification. Figures 1d and 1e show typical current distribution under illumination in individual grains in the channel and above the electrodes, respectively, under a 2 V bias (see the topography and larger scale current mapping for more statistics in Figure S2). The data clearly show that grain boundaries conduct better than the grains in all cases. Moreover, a careful look at one grain in the channel (Figure 1d) shows higher conductivity along straight lines within the grains, which are typical for domain walls,[25] whereas such conductive lines are absent from grains above the electrodes (Figure 1e).

Macroscopic characterization has been performed to demonstrate that the microscale mechanism is also governing the device behavior (Figure 1b). Here, the photocurrent of devices with in-plane and out-of-plane configuration was compared. The measurements indicate the orientational polarization conductance dominancy (in-plane) also at the macroscale.

Complementary microscale and macroscale results with a similar trend for different sample compositions and various biases are given in Figure S3 and S4. The incorporation of 1 and 3 weight % PS into $MAPbI_3$ films increases the photocurrent as compared to plain $MAPbI_3$ (without PS) films, with the 1 weight % PS-$MAPbI_3$ film shows the optimum photocurrent.[26]

Polarization domains are characterized by a memory effect:[27,28] they are switchable by external bias, whereas their orientation is unchanged even upon voltage removal. Hence, if indeed the photocurrent is polarization-domain activated, then we expect to encounter a memory effect. That is if the two sides of a domain are short-circuited, current flows to compensate the polarization-induced surface charge.[11] Figure 1f shows the current distribution under light for a tip-sample short-circuit condition, immediately after the sample was biased with $V_{\text{bias}}$ = 2 V, following Figure 1c. Here, although no bias was applied to the sample during the scan, a solid photocurrent flowed in the after-poling state with a similar distribution as in the biased sample (Figure 1c), albeit with lower values and opposite polarity, as expected (complementary data is given in Figure S5).

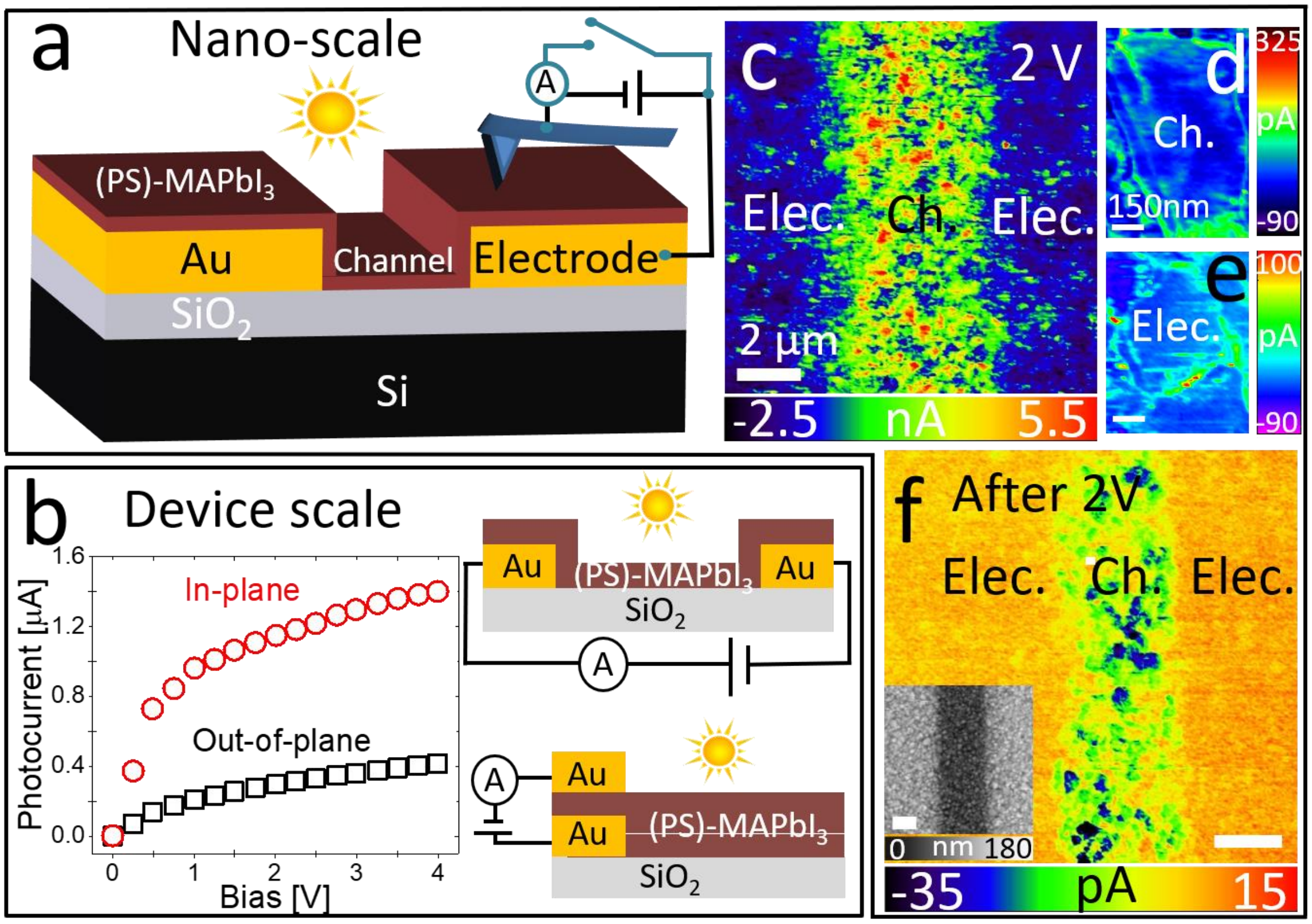

**Figure1| In-plane photocurrent dominancy in an in-plane tetragonal $MAPbI_3$ film with 1 weight % PS**. (**a**) Schematics of the set-up for c-AFM with *in-situ* light and bias. (**b**) Right: sample configuration for in-plane (up) and out-of-plane (bottom) macroscopic current measurements; Left: comparison between out-of-plane and in-plane photocurrent in macroscopic devices, showing that the latter is four times larger than the former. (**c**) c-AFM scan under light and 2 V bias shows that the in-plane channel photocurrent is stronger (up to one order of magnitude) than the out-of-plane current above the electrodes. Current mapping at the sub-micrometer scale of typical individual grains (d) in the channel and (**e**) on electrodes, showing the dominancy of current on grain boundary in both cases and additional conductivity of domain walls in the channel (d). 2 V bias was applied during the scans in (d) and (e). See Figure S2 for the large-scale area current and topography mapping of the areas from which (d) and (e) are taken, as well as

for complementary topography images of the grain and domain walls. (**f**) Current mapping after poling at a $V_{bias}$ of 2 V (see Figure 1a, blue circuit indicating shortcut on the voltage source during the scan), showing a non-volatile memory effect of the photoconductivity. Insert of (f) demonstrates the topography of the scanned area (contact-mode AFM). 200 nm thick 1 weight % PS-$MAPbI_3$ film covers the entire area. Higher regions correspond to the Au electrodes and lower areas are above the $SiO_2$ in the channel.

A prominent fingerprint of polarization domains is their absence upon heating above the tetragonal-to-cubic transition temperature, when the polarization is expected to become null. Consequently, a polarization-activated photocurrent should also vanish above this temperature. Figure 2 shows the temperature dependence of the after-poling photocurrent along with temperature-dependent XRD profiles. The photocurrent map remained nearly unchanged when the temperature was elevated to values as high as 60 °C (Figure 2a-c and Figure S6). Above 60 °C, no photocurrent was measured, in agreement with the phase-transition temperature that was observed in XRD (Figure 2d).[29,30,31] Note that if the photocurrent was due to ion migration and not polarization assisted, the photocurrent should have increased with increasing the temperature ($T$),[32] and not vanish suddenly at the transition temperature: $J = \frac{B\zeta}{T}\exp\left(-\frac{\Delta E}{k_B T}\right)$.[33] Here, $B$ is a constant, $\zeta$ is the electric field, $\Delta E$ is the activation energy for ion migration, and $k_B$ is Boltzmann's constant. Finally, the photocurrent returned to its native distribution when the material was cooled down back to the tetragonal phase (Figure S6).

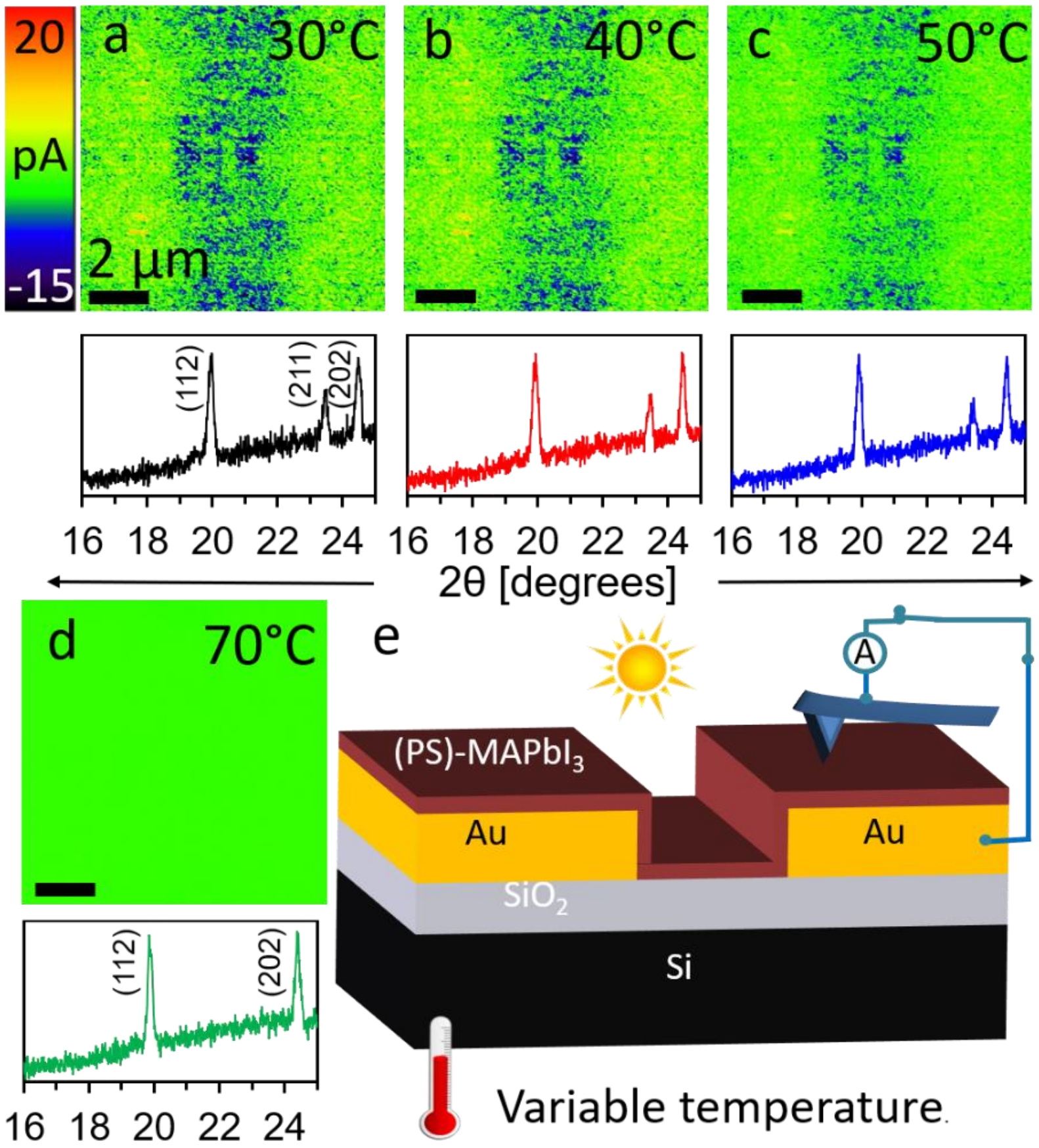

**Figure 2| Disappearance of photocurrent upon heating above the tetragonal-to-cubic transition temperature in 1 weight % PS-MAPbI₃.** c-AFM mapping of the post-poling photocurrent (top) and the XRD profiling (bottom) for (**a**) 30 °C, (**b**) 40 °C, and (**c**) 50 °C showing a slight decrease of the photocurrent with increasing the temperature and no change in the crystal structure**.** (**d**) c-AFM mapping of the post-poling photocurrent (top) and the XRD profiling (bottom) at 70 °C, showing no photocurrent as well as that the crystal structure changed from tetragonal to cubic. The complete temperature-variation cycle that comprises also reoccurrence of the photocurrent at the tetragonal phase after cooling down again is given in Figure S6. (**e**) Schematics of the post-poling variable temperature c-AFM setup.

The photocurrent itself is not enough to indicate device performance for photoconductive-related technologies. Rather, a common figure of merit is the difference in conductance between illumination and dark conditions. Thus, there is a need in revealing the fundamental conductance mechanism not only of the photocurrent but also of dark current. Figure 3a shows the conductivity map of $MAPbI_3$ film, here the scanning was done under dark conditions (see schematics in Figure 3b). As opposed to photoconductivity (Figure 1c), here the dark current flows mainly out-of-plane, above the electrodes (and not in the channel) where the shortest path between the tip and the electrode is. The typical current value is a few tens of pA, in comparison to a few hundreds of pA for photocurrents. Hence, the c-AFM mapping in Figure 3a indicates that as opposed to the photocurrent, the dark current is dominated by ion migration and not by polarization. Likewise, the macroscopic measurements of out-of-plane dark current in Figure 3c show an order of magnitude increase with respect to the measured in-plane dark current. Complementary microscale and macroscale results for different sample compositions and various biases given in Figures S7 and S8 show that PS-$MAPbI_3$ films exhibit reduced dark current as compared to a plain $MAPbI_3$ (without PS) film. This can be attributed to the presence of PS, an insulating material that reduces ion-migration effects.[26]

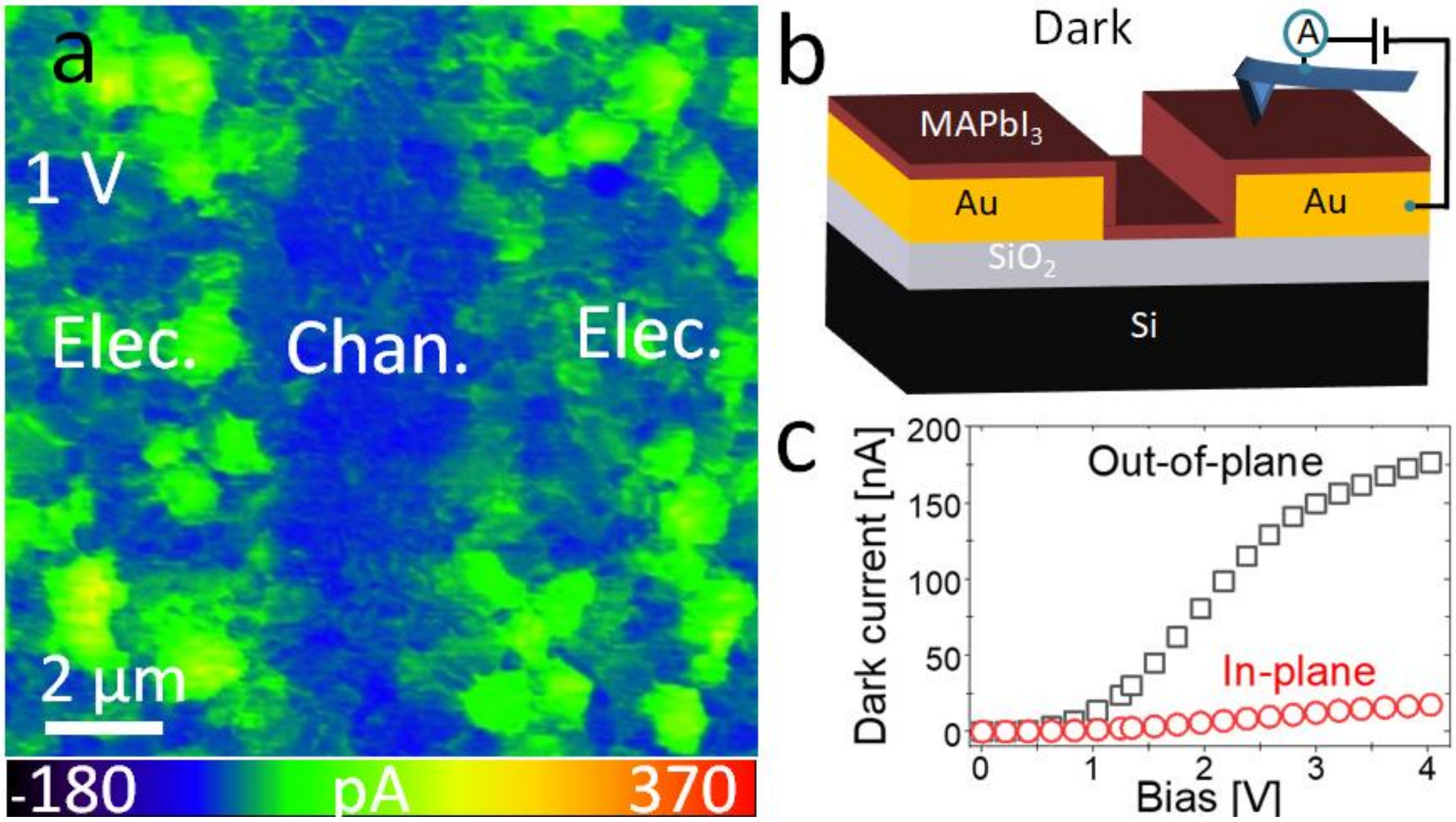

**Figure 3| Out-of-plane dominancy of dark current in an in-plane-polarization $MAPbI_3$ film.** (**a**) c-AFM scans in dark at 1 V of the plain $MAPbI_3$ thin film. Dark current above the electrodes (*i.e.*, out-of-plane current between the electrode and the tip) is higher than the channel current (*i.e.*, in-plane current). The higher currents above the electrode indicate that the dark conductivity is not orientational and follows the shortest path between the tip and the electrode. (**b**) Schematics of the setup for dark c-AFM measurements. (**c**) Comparison between the out-of-plane and in-plane dark current at the macroscopic device scale showing a similar trend as the microscopic measurements and in contrast to the device scale photocurrent under light (see Figure 1b).

## Discussion

The ion migration observed in dark conditions should exist also under illumination. However, because the measured photocurrent is much higher than the dark current, the ion-migration contribution to the photocurrent must be negligible with respect to the polarization-activated photocurrent. The first conclusion from these observations is that light stabilizes the polarization-activated current and perhaps also the polarization domains. This conclusion agrees with earlier observations of photo-stabilized polarization state that occurs *e.g.*, strong charge separation at the domain walls.[25] A second important conclusion is that for improving device performance and the figure of merit, one should enhance the polarization as well as polarization-assisted conductivity and reduce ion migration. Because the importance of ferroelectricity to the photocurrent is known, whereas the intrinsic ferroelectricity of HHPs has been questionable, introducing a polarized ferroelectric polymer, such as polyvinylidene fluoride during processing is a common method to enhance photoconductivity.[34] The above results show that the examined HHP has intrinsic polarization. Hence, adding a non-polar polymer assists in charge segregation between the grains, contributing to photocurrent enhancement. Moreover, a non-polar polymer should decrease the available mobile charge, thus reducing the contribution from ion migration. Therefore, the dark current is expected to decrease, improving the figure of merit. Figures S3, S4, S7, and S8 demonstrate that introducing non-polar polystyrene polymer into $MAPbI_3$ increases the photocurrent and decreases the dark current, both microscopically and macroscopically even for concentrations as low as 1 weight %. Table 1 summarizes the dominant mechanism of conductivity in dark and under light in (PS)-MAPbI3 thin films investigated in this research. The photoconductivity is enhanced in the polarization direction and shows a memristive behavior below the tetragonal to cubic transition temperature. The dark current dominates in out-of-plane and is therefore mainly correlated to the ion migration mechanism.

| Experiments \ Conductivity mechanism | Polarization-assisted | Ion migration | Electron-hole pairing/depairing |
|---|---|---|---|
| Conductivity path: directional or shortest path? | **Directiona**l | Shortest path | Shortest path |
| Dominating conductivity mechanism | **In light** | In dark | In light |
| Non-volatile after-poling photoconductivity | **Yes** | No | No |
| Non-volatile photoconductivity when heating above tetragonal to cubic transition | **No** | No | No |
| Photoconductivity when adding a non-polar polymer to $MAPbI_3$ | **Increases** | Increases | **Increases** |
| Dark conductivity when adding a non-polar polymer to $MAPbI_3$ | **Decreases** | **Decreases** | **Decreases** |

**Table 1| Summary of the expected and observed parameters affecting photoconductive and dark conductivity mechanisms in HHP.** A summary of the expected behavior from polarization-assisted, ion migration and electron-hole pairing/depairing conductivity mechanisms. The experimental observations in this work is highlighted with bold and underline, demonstrating dominancy of polarization-assisted photoconductivity in 1 weight % PS-MAPbI3.

The results in this work demonstrate the polarization-assisted origin of photocurrent in the HHP $MAPbI_3$ materials and devices. The results also indicate the existence of a less dominant ion migration contribution for the photocurrent. The exact origin of polarization in these materials requires additional exploration. A major contribution may arrive from strain-related effects, such as ferroelasticity.[24,25] Flexoelectricity may also play a significant role in the presence of polarization and hence in the observed conductivity.[35] Note that in both ferroelasticity and flexoelectricity, the geometry may affect strongly the strain that originates the polarization. In particular, the thin-film geometry of the examined samples is favorable for enhancing strain-mediated polarization.[36,37] In addition, several works proposed a ferroelectric-based polarization in these materials, which can also support the observations in the current research.[5,9,10,15,38]

Polarization-assisted conductivity is augmented by the presence of light,[17,39] so that in this work, ion migration dominates the dark conductivity in the examined materials. The polarization-assisted conductance dominancy has two major implications about the photoelectric activity of these materials. First, there is a strong temperature dependence, so that properties such as photocurrent are absent above the tetragonal-to-cubic transition temperature. This temperature dependence may serve as the basis for temperature-dependent photoelectric devices. Secondly, because polarization domains are accompanied by a memory effect, at which the switchable polarization is unchanged even in the absence of external voltage, the photoelectric properties of HHP demonstrate a similar memory effect. In particular, the above results show the existence of photocurrent even without bias (post-poling photocurrents in Figure 1f and Figure 2). These results open routes for devices that combine memory and photoelectric activity, with applications ranging from stand-alone switching and power charging devices with remote activation by light to smart solar cells with integrated memristive circuits.

## Methods

***Preparation of $MAPbI_3$ and polystyrene-$MAPbI_3$ precursor solutions*:** $MAPbI_3$ precursor solution was prepared by mixing 0.2305 g of lead iodide ($PbI_2$), 0.0795 g of methylammonium iodide (MAI) in 53.3 µL of dimethyl sulfoxide (DMSO) and 317.5 µL of dimethylformamide (DMF) for 1 h. Further, the 1 and 3 weight % PS-$MAPbI_3$ precursor solutions were prepared by first dissolving the required amount of polystyrene (with a molecular weight of 60,000) in DMSO and DMF. Then, MAI and $PbI_2$ (1/1 by molar) were added to the above PS solution under constant stirring for 1 h (time for crosslinking) at room temperature. The entire process was carried out in ambient conditions.

***Fabrication of HHP thin films:*** For the microscopic conductive measurements, we used silicon chips with 200 nm thermalized $SiO_2$ top layer with an interdigital gold (Au) planar electrodes (2 µm spacing) deposited on top of the oxide layer by photolithography with a Cr adhesion layer. The patterned Au chips were initially washed with Millipore water and later cleaned in an ultra-sonication bath in acetone and isopropanol for 10 minutes each, and finally re-washed with Millipore water. The chips were then treated with piranha for 3 minutes, washed with copious Millipore water, and finally dried with $N_2$ gun. The $MAPbI_3$ or PS-$MAPbI_3$ precursor solution was spin-coated on the cleaned chip at 4000 rpm for 30 seconds. After 6 s of rotation, 200 µL of diethyl ether was dropped onto the rotating chip. Thereafter, the obtained film was annealed at 65 °C for 2 min and 100 °C for 3 min to ensure complete perovskite phase formation. Samples were deposited with a predominant in-plane polarization distribution (Figure S1). Additional films with out-of-plane polarization were deposited between out-of-plane sandwiching electrodes for complementary macroscopic measurements. For the macroscopic conductive measurements, we sputtered 80 nm thick Au on top of the glass ($SiO_2$) substrate in a vacuum chamber using a shadow mask to pattern the Au electrodes.

***Microscopic and macroscopic conductive measurements:*** To characterize dark and photo conductivity at the microscale, we used variable-temperature c-AFM with *in-situ* adjustable illumination and a constant 40% humidity rate (MFP-3D with a customized temperature-control setup, Asylum Research by Oxford Instruments, Inc.). Samples were illuminated with a fiber optic from a remote halogen source. The AFM photodiode zeroing was adjusted to compensate for cantilever heating due to illumination. Voltage was applied *in situ* between the AFM tip and the lateral electrodes ($V_{bias}$), allowing both in-plane and out-of-plane conductivity characterization for areas between (in the channel) and above the electrodes, respectively (see Figure 1a). Note that conductivity profiling was done both during the biasing (poling) as well as post-poling, with no voltage application.

Macroscopic photoconductive measurements were performed with a solar simulator (see Figure 1b, right) in two configurations: (i) in-plane, when lateral bias was applied across the $MAPbI_3$ channel between the two sets of adjacent interdigital Au electrode fingers; (ii) out-of-plane when the bias is applied to $MAPbI_3$ film sandwiched between two gold electrodes. For the macroscopic conductive measurements, a two-probe method was used by connecting one probe to one gold electrode on a glass substrate and another probe being connected to the other gold electrode either on glass or on the sample (based on the configuration). The distance between the Au electrodes in the out-of-plane and in-plane configuration is 200 nm and 2 µm, respectively, similar to the microscopic measurements. The current-voltage characteristics of the Au/(PS)-$MAPbI_3$/Au devices were measured in dark and under the AM 1.5 G (one sun, 100 mW $cm^{-2}$) illumination using a Keysight 3458A Digital multimeter. The AM 1.5 G light was generated using a Xenon-lamp based solar simulator (Newport Oriel Instrument 67005, 150 W Solar Simulator), and the light intensity was calibrated by a NREL calibrated KG5 silicon reference cell. All measurements were conducted in ambient air with 30-40% relative humidity and at 20 °C.

**Structural characterization:**

XRD measurements were performed by following the $\theta - 2\theta$ method in a Rigaku SmartLab diffractometer with a Cu electrode ($\lambda$ = 0.15406 nm). Temperature-dependent measurements were performed under nitrogen atmosphere and a dedicated heating sample stage was used. The thin-films were prepared on a glass substrate for the XRD measurements.

## Acknowledgments

The researchers would like to thank the Technion-Waterloo foundation for funding the project. The Technion team acknowledges support from the Zuckerman STEM Leadership Program and the Technion Russel Barry Nanoscience Institute, as well as from the Israel Science Foundation (ISF) Grant No. 1602/17. We would also like to thank Dr. Maria Koifman for helping with the XRD measurements and Dr. Asaf Hershkovitz and F. Ted Limpoco from Asylum Research by Oxford Instruments for fruitful discussions.

## Author contributions

R. S. and C. S. contributed equally to this work. C. S. and R. S. performed the c-AFM measurements and Y. I. joined them for the data analysis. R. S. prepared the samples, as well as made the macroscopic device measurements and interpreted them. H. E. carried out the XRD measurements at variable temperatures. All authors participated to the preparation of the manuscript. Y. I. and V. M. conceived and directed this project, whereas V. M. supervised the sample preparation macroscale and Y. I. supervised the AFM and XRD characterizations.

# Polarization Origin of Photoconductivity in $MAPbI_3$ Thin Films

Rohit Saraf,[1] Cecile Saguy,[2] Vivek Maheshwari,[1] Hemaprabha Elangovan[2,3] and Yachin Ivry[2,3,*]

[1]Department of Chemistry, Waterloo Institute of Nanotechnology, University of Waterloo, Waterloo, Ontario N2L 3G1, Canada

[2] Solid State Institute, Technion – Israel Institute of Technology, Haifa 32000, Israel

[3]Department of Material Sciences and Engineering, Technion – Israel Institute of Technology, Haifa 32000, Israel

## Supplementary information

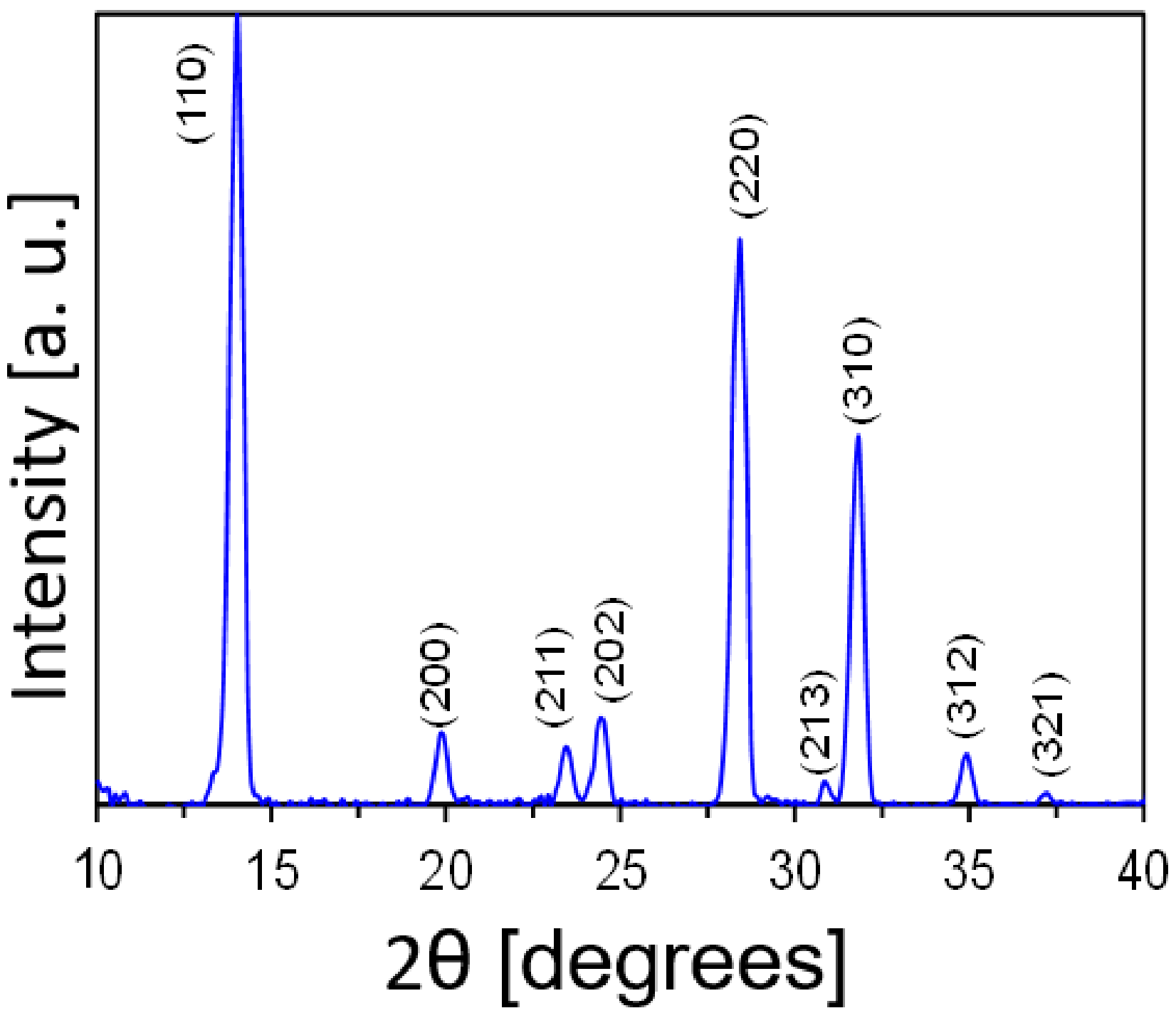

**Figure S1| Dominant in-plane tetragonal structure of the thin $MAPbI_3$ films**. Indexed XRD pattern that corresponds to the tetragonal structure in a highly crystalline $MAPbI_3$ film. In-plane and out-of-plane polarization are estimated based on the intensity ratio of the respective peaks. The intensity ratio calculated between the (202) and (220) peaks is $\frac{I_{(202)}}{I_{(220)}} = 0.15$, in comparison to the 0.28 standard intensity ratio in JCPDS, (number: 01-085-5508) showing a predominant in-plane polarization.

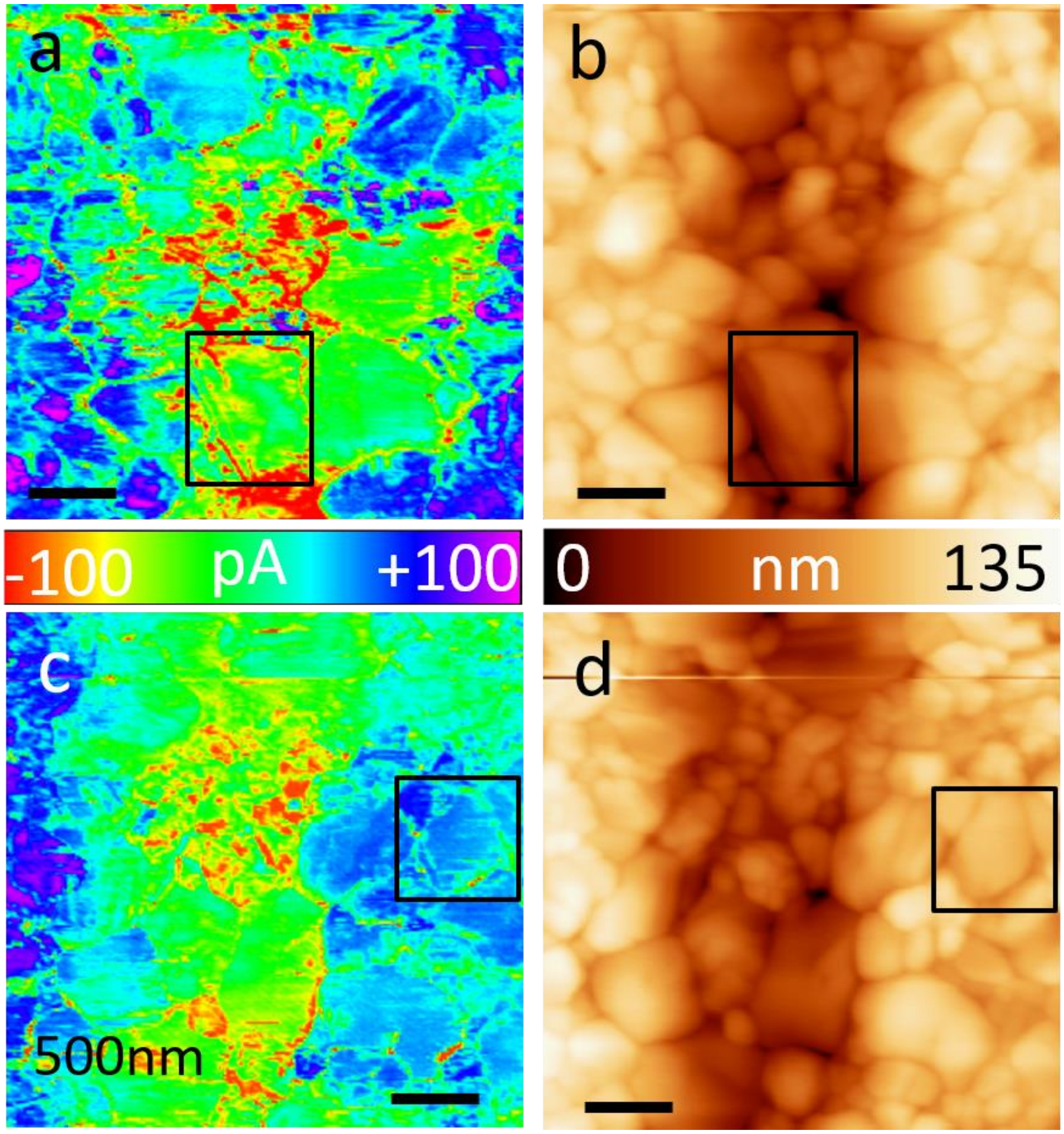

**Figure S2| Photocurrent and topography mapping of individual grains at the sub-microscopic scale.** Black squares highlight conductivity around typical individual grains (**a and b**) in channel, and (**c and d**) on electrodes, corresponding to the current mapping displayed in Figure 1d and 1e. Domain walls appear as straight lines within the grain in (a-b).

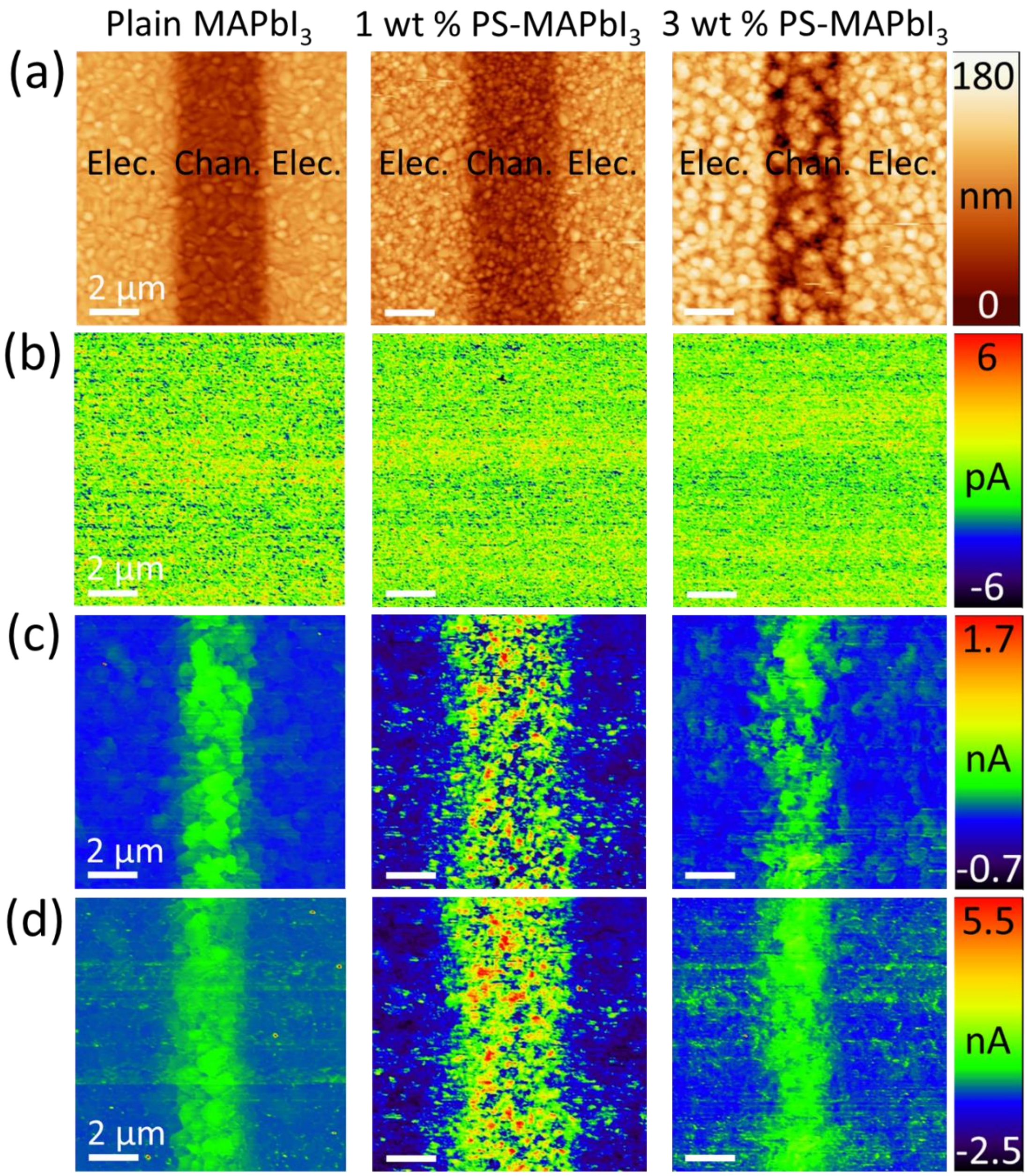

**Figure S3| In-plane photocurrent dominancy in the in-plane-polarization $MAPbI_3$ films**. (**a**) Topography of the scanned area (contact-mode AFM) of the plain (0), 1, and 3 weight % polystyrene-$MAPbI_3$ films. High regions are above the Au electrodes and low areas are above the $SiO_2$ in the channel. Current mapping images under light at bias of (**b**) 0 V, (**c**) 1 V and (**d**) 2 V showing that the in-plane channel photocurrent is stronger than the out-of-plane current (above

the electrodes) for both plain and polystyrene-$MAPbI_3$ films. The data show clearly that adding 1 weight % polystyrene to the polystyrene-$MAPbI_3$ films enhances the effect.

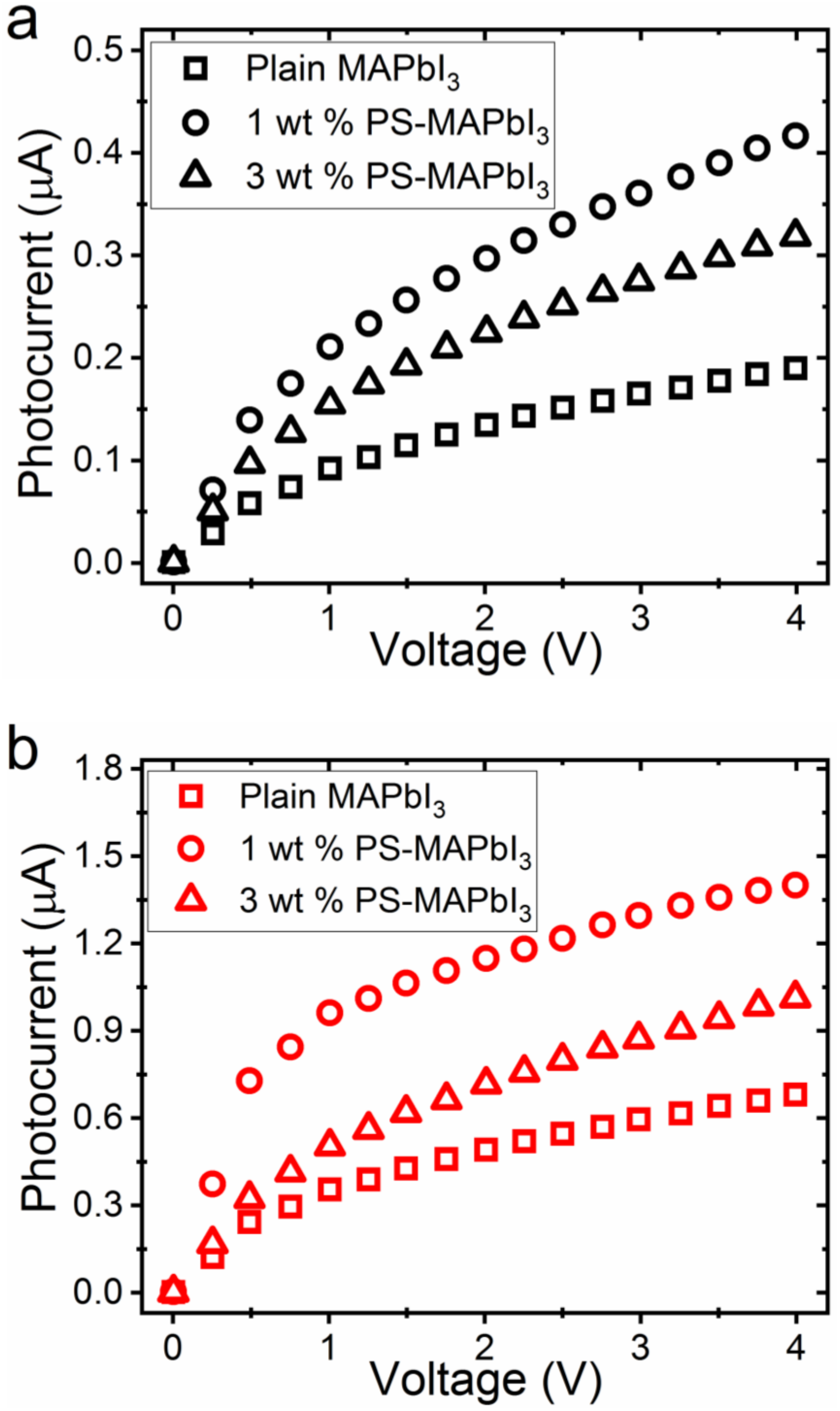

**Figure S4| In-plane macroscopic photocurrent dominancy in the in-plane-polarization $MAPbI_3$ films**. Macroscopic photocurrent measurements of $MAPbI_3$ films with 0 (plain), 1, and 3 weight

% polystyrene in (**a**) out-of-plane and (**b**) in-plane configurations as a function of the applied voltage. In both configurations, the incorporation of polystyrene into $MAPbI_3$ film increases the photocurrent as compared to plain $MAPbI_3$ (without polystyrene) film. The 1 weight % polystyrene-$MAPbI_3$ film shows the optimum photocurrent. The in-plane photocurrent is almost four times larger than the out-of-plane photocurrent for all 3 samples.

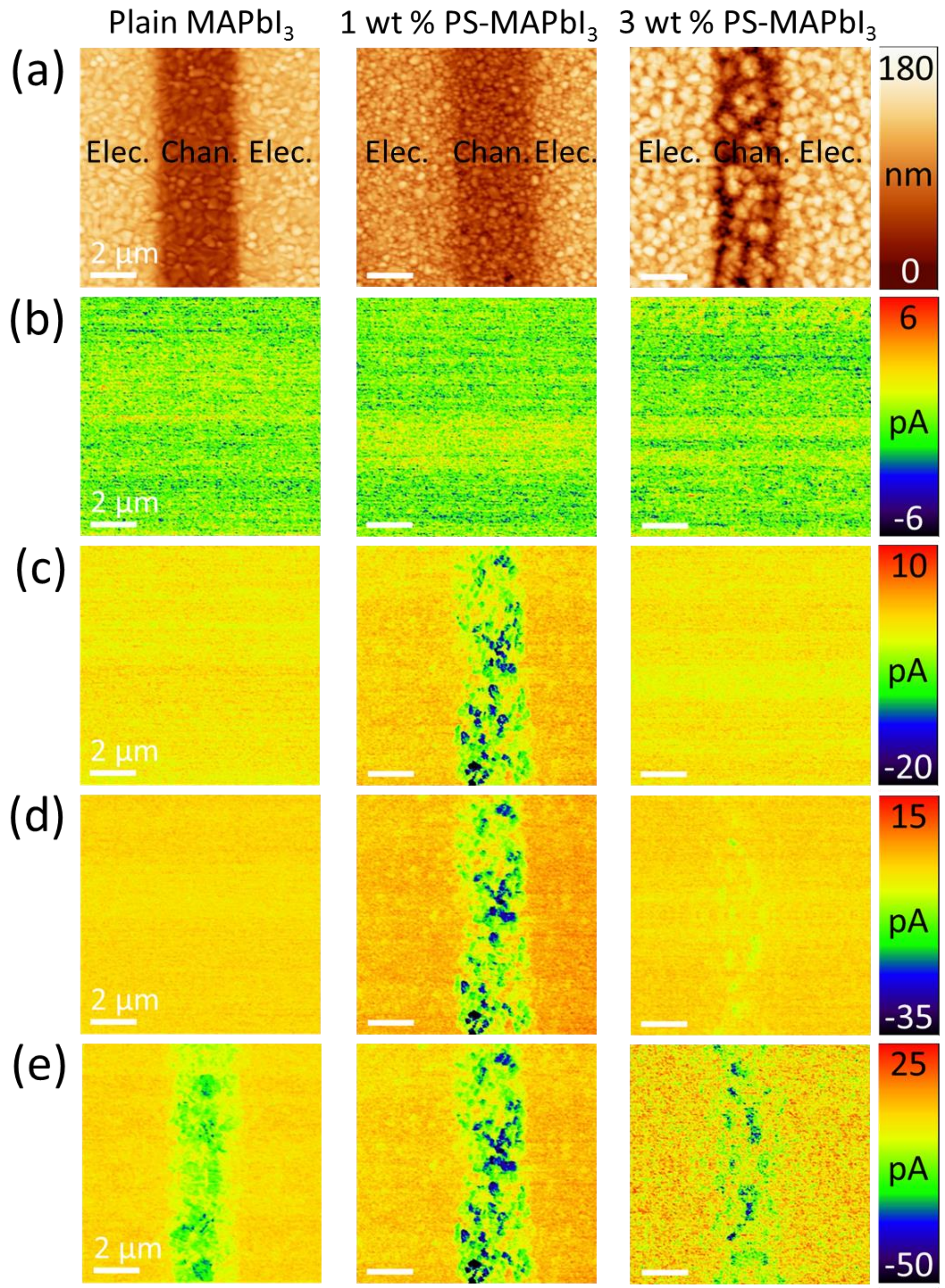

**Figure S5| Memristive photoconductivity in $MAPbI_3$ films measured under light during non-volatile after-poling scan under light.** Memristive photoconductivity in the presence of light showing the dominancy of the polarization effects in $MAPbI_3$ films that were pre-poled. (**a**) Topography, and (**b-e**) current mapping images after poling the samples at a $V_{bias}$ of (**b**) 0 V, (**c**) 1

V, (**d**) 2 V, and (**e**) 4 V under light showing a non-volatile memory effect of the photoconductivity. As the 1 weight % polystyrene-$MAPbI_3$ sample shows a higher photocurrent in the channel as compared to plain $MAPbI_3$ and 3 weight % polystyrene-$MAPbI_3$ samples (Figure S3), therefore it leads to greater internal electric field strength that results in higher ferroelectric or memory effect.

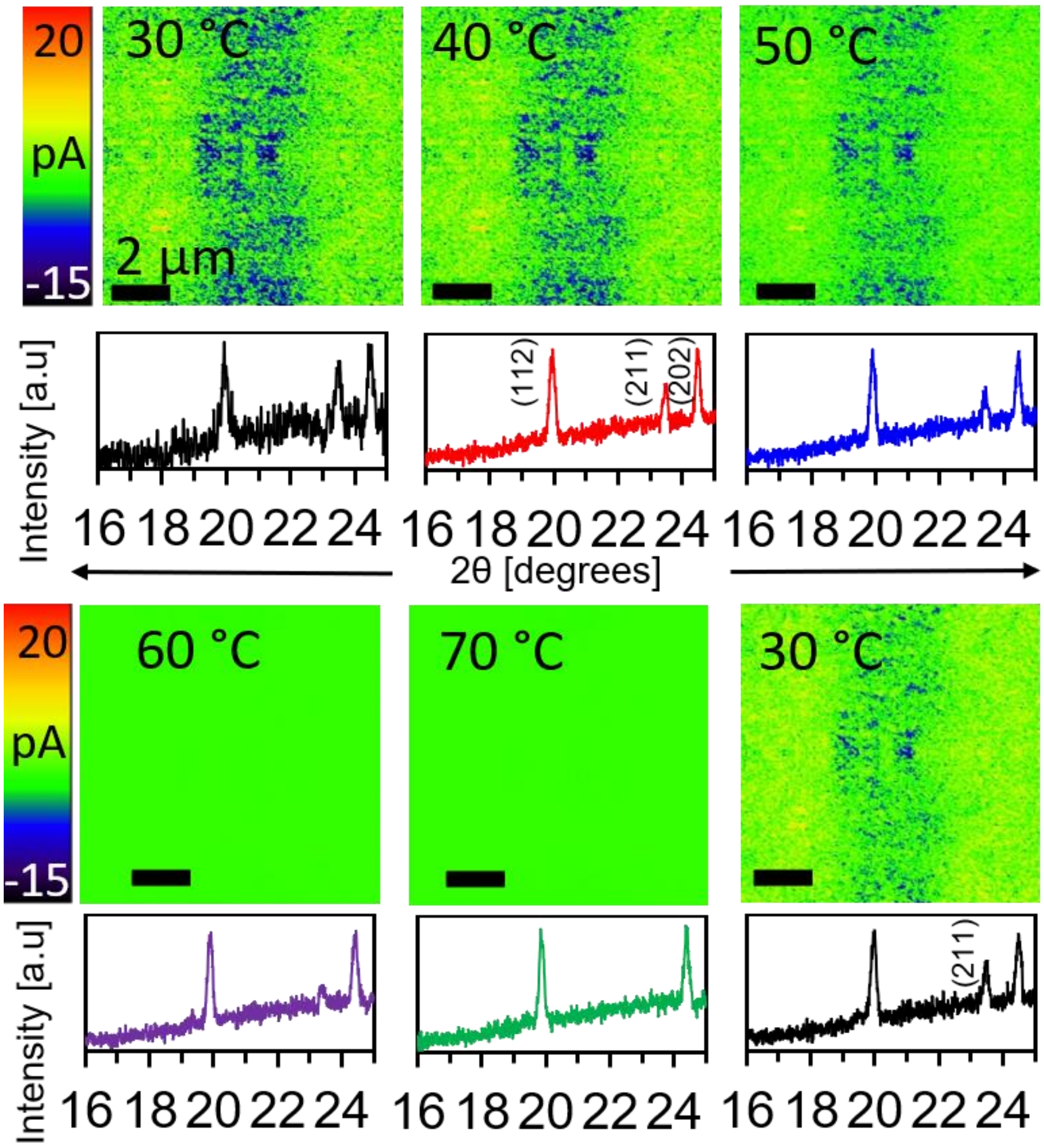

**Figure S6| Disappearance of photocurrent upon heating above the tetragonal-to-cubic transition temperature (57 °C) in 1 weight % polystyrene-$MAPbI_3$ and reoccurrence of the photocurrent at the tetragonal phase after cooling down again from 70 °C to 30 °C**. c-AFM

mapping of the post-poling photocurrent (top) and the corresponding XRD profiling (bottom) during a complete temperature cycle showing photoconductivity at the tetragonal structure, no conductivity when the material transfers to cubic, and reoccurrence of the photoconductivity when the material is cooled down back to the tetragonal structure. Note that the slight inconsistency between the c-AFM and the XRD at 60 °C is due to temperature calibration difference between the two systems.

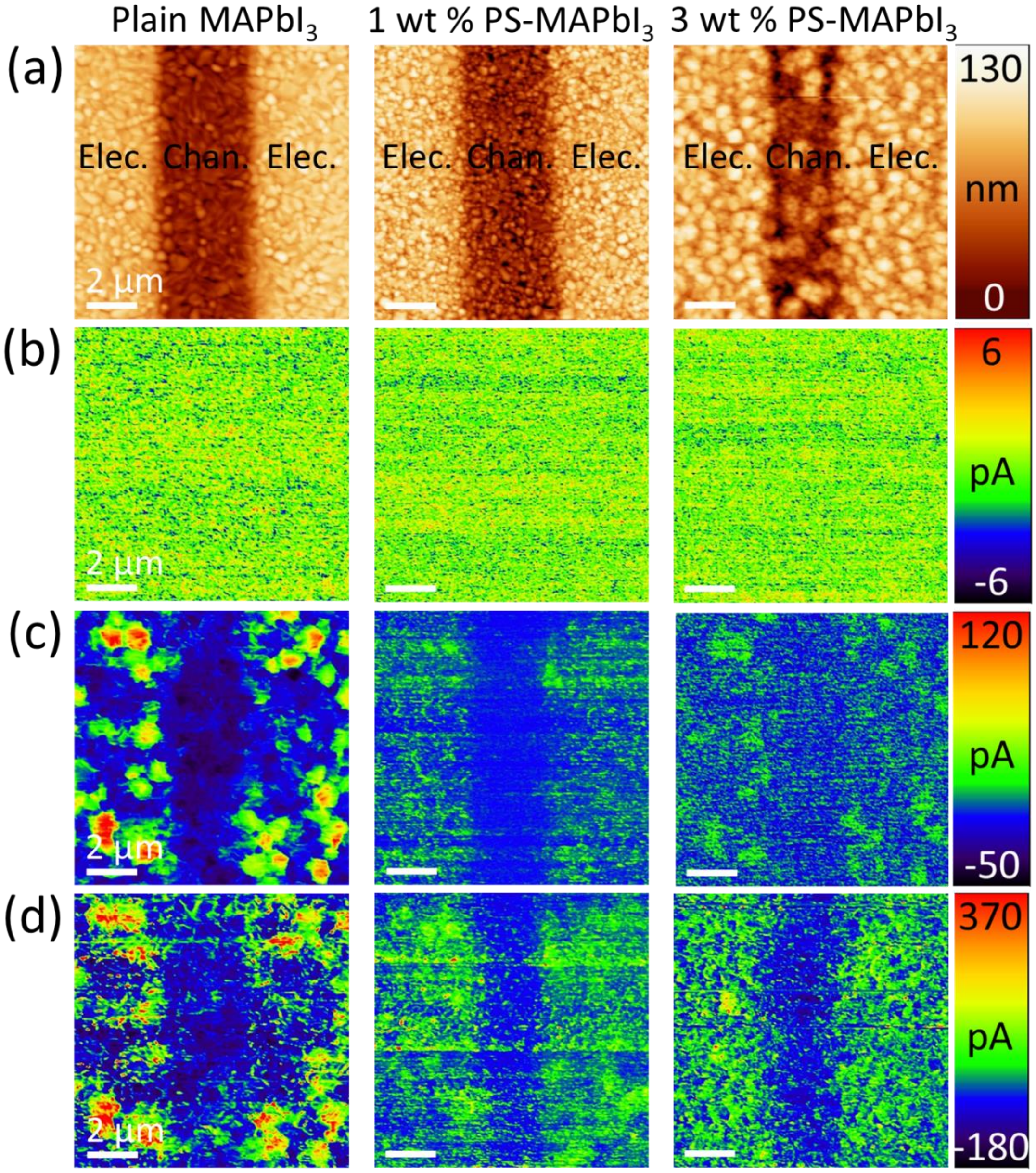

**Figure S7| Out-of-plane dark current dominancy in the in-plane-polarization $MAPbI_3$ films**. (**a**) Topography of the scanned area of the plain (0), as well as 1, and 3 weight % polystyrene-$MAPbI_3$ films. High regions are above the Au electrodes and low areas are above the $SiO_2$ in the channel. (**b-d**) Current mapping images in dark condition at a bias of (**b**) 0 V, (**c**) 1 V, and (**d**) 2 V showing that the out-of-plane dark current above the Au electrodes is higher than the in-plane dark current in the channel. The higher currents above the electrode indicate that the dark conductivity is not orientational and follows the shortest path between the tip and the electrode as expected from ion mobility and not from polarization-assisted conductivity.

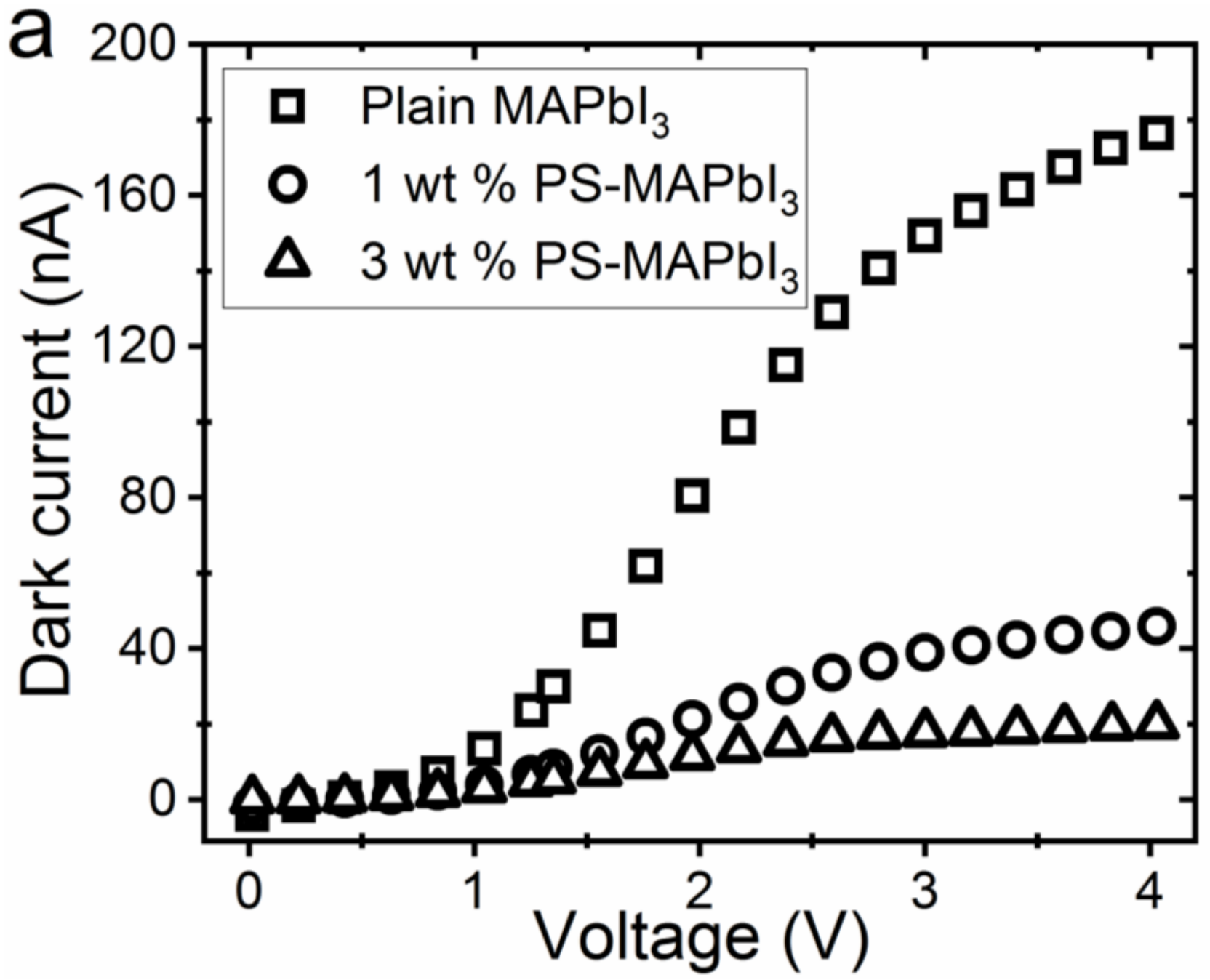

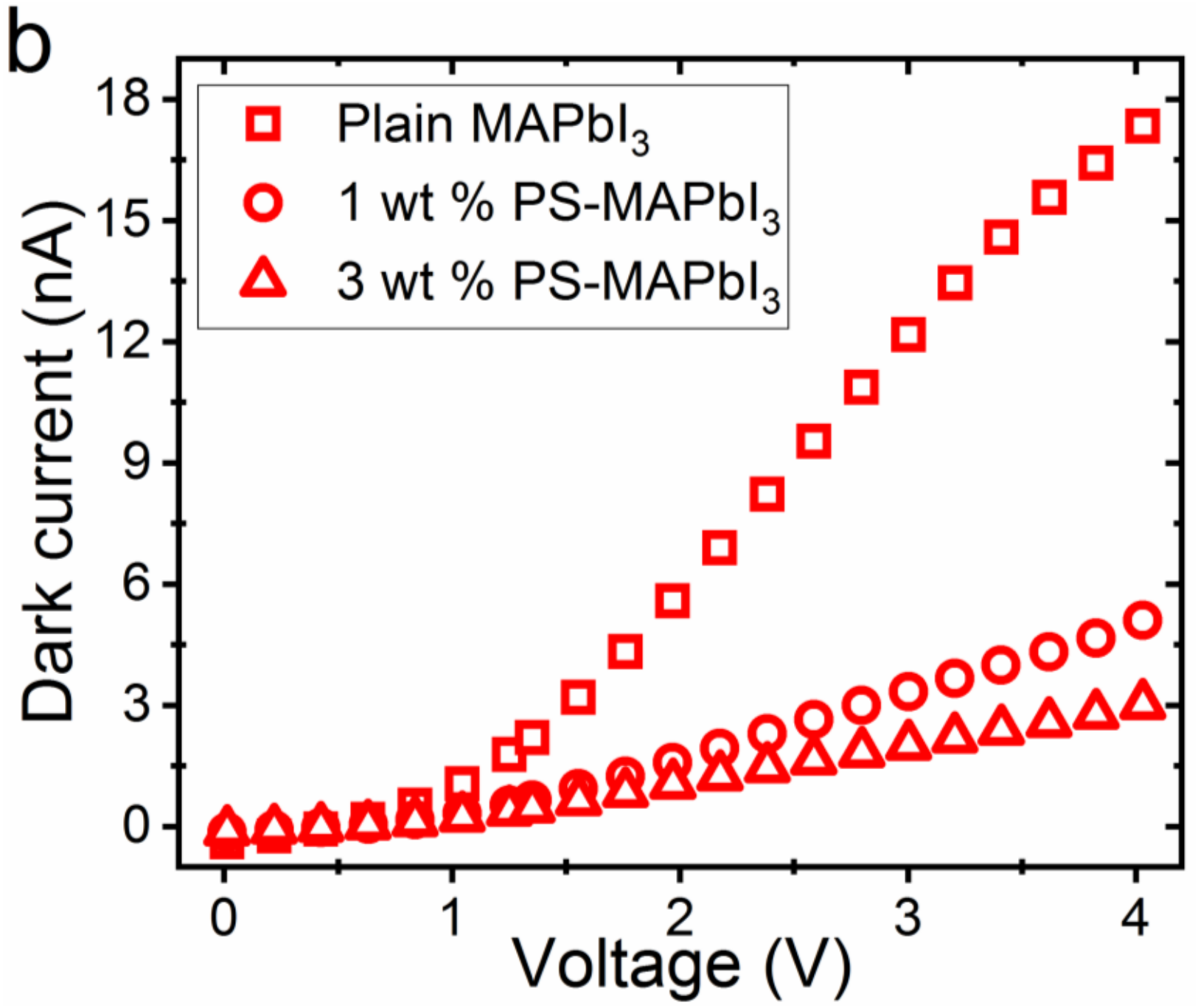

**Figure S8| Dominancy of the out-of-plane macroscopic dark current in the in-plane-polarization $MAPbI_3$ films**. Macroscopic dark current measurements of $MAPbI_3$ films with 0 (plain), 1, and 3 weight % polystyrene in (**a**) out-of-plane and (**b**) in-plane configurations as a function of the applied voltage. In both configurations, the incorporation of polystyrene into $MAPbI_3$ film reduces the dark current significantly in comparison to plain $MAPbI_3$ (without polystyrene) film due to reduced ion-migration effect. The out-of-plane dark current is almost 10 times higher than the in-plane dark current for all 3 samples.